\pdfoutput=1
\documentclass[prb,twocolumn,superscriptaddress,citeautoscript,showpacs]{revtex4-1}
\usepackage{mmap}
\usepackage{graphicx}
\usepackage{longtable}
\usepackage{amsmath}
\usepackage{amstext}
\usepackage{amssymb}
\usepackage{amsfonts}
\usepackage{color}
\usepackage{dcolumn}
\usepackage{bm}
\usepackage{dsfont}
\usepackage[breaklinks=true,colorlinks=true,linkcolor=blue,urlcolor=blue,citecolor=blue]{hyperref}

\newcommand{\ri}{{\rm i}}
\newcommand{\ra}{\rangle}
\newcommand{\la}{\langle}

\newcommand{\vk}{{\mathbf{k}}}
\newcommand{\vq}{{\mathbf{q}}}

\newcommand{\ut}[1]{\mathrm{\; #1}}
\newcommand{\sgn}{\operatorname{sgn}}
\newcommand{\Tr}{\operatorname{Tr}}

\renewcommand{\Im}{\operatorname{Im}}
\renewcommand{\Re}{\operatorname{Re}}
\newcommand{\STO}{SrTiO$_3$}
\newcommand{\BTO}{BaTiO$_3$}
\newcommand{\ep}{$e$-ph}

\begin{document}

\title{Phonon linewidth due to electron-phonon interactions with strong forward scattering in FeSe thin films on oxide
substrates}

\author{Yan~Wang}
\affiliation{Department of Physics and Astronomy, University of Tennessee, Knoxville, Tennessee 37996, USA}

\author{Louk~Rademaker}
\affiliation{Kavli Institute for Theoretical Physics, University of California Santa Barbara, California 93106, USA}

\author{Elbio~Dagotto}
\affiliation{Department of Physics and Astronomy, University of Tennessee, Knoxville, Tennessee 37996, USA}
\affiliation{Materials Science and Technology Division, Oak Ridge National Laboratory, Oak Ridge, Tennessee 37831, USA}

\author{Steven~Johnston}
\affiliation{Department of Physics and Astronomy, University of Tennessee, Knoxville, Tennessee 37996, USA}

\date{\today}

\begin{abstract}
The discovery of an enhanced superconducting transition temperature $T_c$ in monolayers of FeSe grown on
several oxide substrates has opened a new route to high-$T_c$ superconductivity through interface
engineering. One proposal for the origin of the observed enhancement is an electron-phonon ({\ep})
interaction across the interface that peaked at small momentum transfers. In this paper, we examine the
implications of such a coupling on the phononic properties of the system. We show that a strong forward
scattering leads to a sizable broadening of phonon lineshape, which may result in charge instabilities at
long-wavelengths. However, we further find that the inclusion of Coulombic screening significantly reduces
the phonon broadening. Our results show that one might not expect anomalously broad phonon linewidths in
the FeSe interface systems, despite the fact that the {\ep} interaction has a strong peak in the forward
scattering (small $\vq$) direction.
\end{abstract}


\maketitle

\section{Introduction}
Due to its structural simplicity, FeSe has played a leading role in many experimental and theoretical studies
on Fe-based superconductors since its discovery in 2008~\cite{Hsu2008}. The enduring interest in this
compound is partially owed to the high $T_c$ (ranging between $55\text{--}100\ut{K}$) achieved when monolayer
FeSe films are grown on {\STO} substrates~\cite{Wang2012a,Ge2015} (FeSe/STO), a ten-fold enhancement from the
$T_c\sim 8\ut{K}$ of bulk FeSe crystals at ambient pressure~\cite{Hsu2008}. Intriguingly, the high $T_c$ in
the interfacial system proves to be robust for various oxide substrates, including {\STO}
(001)~\cite{Wang2012a,Lee2014}, {\BTO} (001)~\cite{Peng2014a}, {\STO}
(110)~\cite{Zhang2016b,Zhou2016a,Zhang2016a}, anatase TiO$_2$ (001)~\cite{Ding2016}, and rutile TiO$_2$
(100)~\cite{Rebec2016}. These oxide substrates, terminated at TiO$_2$ surface when interfaced with FeSe, have
lattice parameters significantly larger than that of bulk FeSe and thus apply strong tensile strain on FeSe
thin films. The anatase and rutile TiO$_2$ substrates even induce rather different strains along $a$ and $b$
axes of the monolayer FeSe. The $T_c$'s, however, are consistently above $55\ut{K}$, as measured by
angle-resolved photoemission spectroscopy (ARPES). This observation appears to rule out a direct correlation
between the enhanced superconductivity and the tensile strain~\cite{Peng2014a,Rebec2016}.

The electronic structure of the interfaces displaying enhanced $T_c$'s are also remarkably similar across the
various substrates. For instance, the Fermi surface measured by ARPES consists of only electron pockets at
the corners of the two-Fe Brillouin zone, indicating substantial electron doping from the parent compound.
This observation poses a challenge to theories for the high $T_c$ based on the pairing mediated by spin
fluctuations that are strongly enhanced by Fermi surface nesting. One potential solution to this problem is
the involvement of bands below the Fermi level in pairing (so-called incipient band
pairing)~\cite{Bang2014,Chen2015,Linscheid2016,Mishra2016a}. Another possibility is the involvement of a
different type of pairing mediator such as nematic fluctuations~\cite{Kang2016} or phonons particular to the
interface~\cite{Coh2015,Lee2014,Wang2012a}. Evidence for the latter has been provided by the common
observation of replica bands in the electronic structure of superconducting FeSe monolayers on {\STO}
\cite{Lee2014,Zhang2016a}, {\BTO}~\cite{Peng2014a}, and rutile TiO$_2$~\cite{Rebec2016}.

The replica bands observed by ARPES are exact copies of the original bands crossing the Fermi level in
momentum space but with a weaker spectral weight. They are interpreted as being generated by an
electron-phonon ({\ep}) interaction between the FeSe electrons and oxygen phonons in the substrate
\cite{Lee2014, Lee2015,Rademaker2016}. This view is supported by the fact that the $\sim 100\ut{meV}$ energy
offset between the primary and the replica band coincides with the phonon energy of oxygen modes in
{\STO}~\cite{Li2014}, {\BTO}~\cite{yWang2016a}, and TiO$_2$~\cite{Rebec2016}. Due to the particular
properties of the interface~\cite{Lee2015,Kulic2017,Zhou2016}, this interaction is strongly peaked for
forward scattering (i.e., peaked at small momentum $|\vq|$ transfer), as found by analyzing the electrostatic
potential from the dipole induced by the oxygen modes~\cite{Lee2014,Lee2015,Kulic2017} and by
first-principles calculations~\cite{yWang2016a,Li2014,Zhou2016}. This unique momentum structure accounts for
the fact that the replicas sharply trace the dispersion of the primary band, which requires the {\ep}
interactions be forward-focused. Such a coupling can also significantly enhance $T_c$, due to the linear
dependence of $T_c$ on the dimensionless coupling constant $\lambda_m$~\cite{Rademaker2016,Kulic2017}, as
opposed to the exponential dependence obtained for the usual BCS case. For example, assuming a narrow width
$q_0$ for the forward scattering peak, some of the current authors found $\lambda_m\sim 0.15\text{--}0.2$
reproduces the measured spectral weight ratio between the replica band and the primary band and at the same
time a $T_c\sim 60\text{--}70\ut{K}$~\cite{Rademaker2016}. Ref.~\onlinecite{Aperis2017} has obtained similar
results after extending this approach to a more realistic band structure.

Many aspects of the influence of the {\ep} interactions with strong forward scattering on electronic
properties and superconductivity are summarized in Refs.~\onlinecite{Rademaker2016,yWang2016,Kulic2017}. In
comparison, there are no qualitative or quantitative studies of the phononic properties for the problem at
hand. Here, we have carried out such a study to address two issues. First, Zhang \textit{et
al.}~\cite{Zhang2016} recently measured the phonon linewidth of a $\sim 90\ut{meV}$ phonon mode penetrating
from the {\STO} substrate into thin FeSe Films using high resolution electron energy loss spectroscopy
(HREELS) and concluded a mode-specific {\ep} coupling constant $\lambda\sim 0.25$. Not only does this echo
the discovery of replica bands by the ARPES experiments in the same system, but it also calls for a
theoretical consideration on the HREELS measurements. Doing so would corroborate both the total coupling
strength and momentum dependence of the {\ep} coupling in FeSe/STO system with those inferred from the ARPES
measurements. Second, when a strong {\ep} coupling is distributed over a subset of wave vectors, one expects
tendencies towards charge-density-wave formation that can compete with superconductivity. Such tendencies
will manifest themselves as Kohn anomalies in the phonon dispersion and broad phonon linewidths. One can,
therefore, address this issue directly by examining the phononic self-energy.

Here, we examine the phonon linewidth due to {\ep} interactions with strong forward scattering using the same
model adopted in Ref.~\onlinecite{Rademaker2016} to study the electronic spectral function. We first describe
the details of the model and method in Sec.~\ref{sec:model}. Next, in Sec.~\ref{sec:analytical} we give some
analytical results for the normal state phonon properties in the \emph{perfect} forward scattering limit,
where the interaction is treated as a delta function at ${\bf q} = 0$. Our numerical results for both normal
and superconducting state with finite $q_0$ are given in Sec.~\ref{sec:numerical}. Here, our results show
that the forward focused peak in the {\ep} coupling results in very broad phonon lineshapes. However, in
Sec.~\ref{sec:screened_linewidth} we reintroduce Coulomb screening, which subsequently undresses the phonon
propagator and suppresses these effects. Finally, in Sec.~\ref{sec:summary} we summarize our results and make
some concluding remarks in relation to the HREELS experiment of Zhang \textit{et al.}~\cite{Zhang2016}

\section{Model and Method} \label{sec:model}
Our model Hamiltonian describes a single band model of FeSe electrons coupled to an optical phonon branch via
a momentum-dependent coupling, which reads
\begin{align} \label{eq:elphHam}
  H =
  & \sum_{\vk,\sigma} \xi^{\phantom\dag}_\vk c^\dag_{\vk,\sigma}c^{\phantom\dag}_{\vk,\sigma}
    + \sum_{\vq} \omega^{\phantom\dag}_\vq \left( b^\dag_{\vq}b^{\phantom\dag}_\vq + \frac{1}{2}\right) \notag \\
  & + \frac{1}{\sqrt{N}}\sum_{\vk,\vq,\sigma} g(\vk,\vq)c^\dag_{\vk+\vq,\sigma}
    c^{\phantom\dag}_{\vk,\sigma} (b^\dag_{-\vq} + b^{\phantom\dag}_\vq).
\end{align}
Here, $c^\dag_{\vk,\sigma}$ ($c^{\phantom\dag}_{\vk,\sigma}$) creates (annihilates) an electron with
wavevector $\vk$ and spin $\sigma$, $b^\dag_\vq$ ($b^{\phantom\dag}_\vq$) creates (annihilates) a phonon with
wavevector $\vq$; $\xi_\vk$ is the electronic band dispersion measured relative to the chemical potential
$\mu$; $\omega_\vq$ is the phonon dispersion ($\hbar = 1$); and $g(\vk,\vq)$ is the momentum dependent {\ep}
coupling.

We take a simple electronic band dispersion $\xi_\vk = -2t[\cos(k_x a)+\cos(k_y a)] - \mu$, where $a$ is the
in-plane lattice constant. We set $t = 0.075\ut{eV}$ and $\mu = -0.235\ut{eV}$, which produces around
$\Gamma$ point an electronlike Fermi pocket with $k_\text{F} = 0.97/a$, a Fermi velocity $v_\text{F} =
0.12\ut{eV}\cdot a/\hbar$ along the $k_y = 0$ line, and an effective electron band mass $m^{*}_{x,y} = \left(
\frac{\partial^2 \xi_\vk}{\hbar^2\partial k_{x,y}^2} \right)_{\vk=0}^{-1} = \frac{\hbar^2}{2ta^2} = 3.3 m_e$,
which is similar to the electron pocket at the $M$ point in FeSe/STO seen in ARPES
experiments~\cite{Liu2012,He2013,Lee2014}. Since we have a single band model, it only takes a trivial
$\mathbf{Q}=(\pi/a, \pi/a)$ shift to map our $\Gamma$-point pocket onto the electron pocket in the real
system centered at the $M$ point and any physical quantities depending only on the momentum transfer $\vq =
\vk-\vk'$, such as phonon linewidth, do not depend on the position of the pocket. Since we are not
considering the effects of an unconventional pairing mechanism here, we do not need to consider the
possibility of $d$-wave instabilities due to scattering between the electron pockets. As such, a single band
model is sufficient for our purpose.

Throughout we approximate the experimental phonon dispersion with a dispersionless Einstein mode $\omega_\vq
\approx \omega_\text{ph} = 100 \ut{meV}$ according to the observed energy separation between the replica band
and the primary band~\cite{Lee2014,Rebec2016}, as well as the phonon dispersion of the interface, as measured
by HREELS~\cite{Zhang2016}. We neglect the fermion momentum dependence in the coupling $g(\vk,\vq)=g(\vq)$,
where $\vq$ is the momentum transfer and adopt $g(\vq)=g_0\sqrt{8\pi/(aq_0)^2}\exp(-|\vq|/q_0)$ as derived
from simple microscopic model~\cite{Lee2014,Lee2015,Kulic2017}. Here, $g_0$ is adjusted to fix the total
dimensionless coupling strength of the interaction and $q_0$ sets the range of the interaction in momentum
space. The normalization factor $\sqrt{8\pi/(aq_0)^2}$ is chosen such that $\la g^2(\vq) \ra_\vq \approx
g_0^2$ for $q_0\ll 2\pi$, where $\la F_\vq \ra_\vq = a^2\iint_\text{BZ} F_\vq dq_x dq_y /(2\pi)^2$ denotes an
momentum integral over the first Brillouin zone. We will typically set the in-plane lattice constant $a = 1$
below; however, we will occasionally write it out for clarity.

The values of $\omega_\vq$ and $g(\vq)$ we use in the calculation include all the screening effect within the
oxide substrate, but \emph{none} from the FeSe film. Thus, we refer to them as the ``bare'' or ``unscreened''
quantities. In Sec.~\ref{sec:screened_linewidth}, we show that such a treatment is justified in calculating
the electron self-energy using the ``unscreened'' phonon propagator \emph{and} the ``unscreened'' coupling
$g(\vq)$, but it overestimates the phonon self-energy, especially the imaginary part (phonon linewidth) at
$\vq=0$, by overlooking the strong screening effect at $\vq=0$ from the FeSe film. The difference between the
fully screened phonon frequency $\omega_\text{ph}$ (by both the substrate and the FeSe film) and partially
screened $\omega_\vq$ (only by the substrate itself) is small, however, so we do not distinguish them
($\omega_\vq \approx \omega_\text{ph}$) in sections \ref{sec:model}, \ref{sec:analytical}, and
\ref{sec:numerical}. Our calculation in Sec.~\ref{sec:screened_linewidth} shows that the difference is within
10\% for most parameters. The experimental measurements in Ref.~\onlinecite{Zhang2016} on phonon frequency in
{\STO} with and without FeSe deposited also support this conclusion.

\begin{figure}
  \centering
  \includegraphics[width=1\columnwidth]{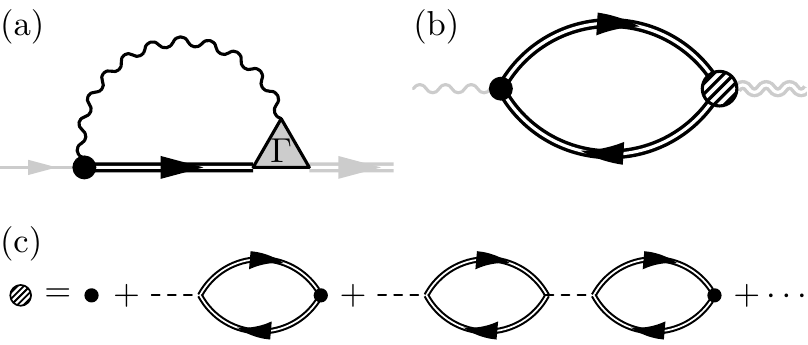}

\caption{The Feynman diagram for the electron self-energy (a) and the phonon self-energy (b). The extra
external legs (gray lines) are not part of self-energy but are attached for clarity. The lines (double-lines)
with an arrow in the middle represent bare (dressed) electron propagators; the wiggly-lines
(double-wiggly-lines) represent bare (dressed) phonon propagators. The gray triangle represents the vertex
part. (c) The screened electron-phonon vertex, approximated by a series involving Coulomb interactions
(dashed lines) and neglecting vertex corrections from the crossing diagrams.}
  \label{fig:diagrams}
\end{figure}

The electron and phonon self-energies due to {\ep} interaction are calculated using Migdal-Eliashberg theory,
where the vertex part $\Gamma(\ri\omega_n,\vk;\ri\omega_\nu,\vq)$ is approximated with the zeroth order
vertex function $g(\vq)$. Here, $\omega_n$ ($\omega_\nu$) is the fermionic (bosonic) Matsubara frequency.
This is shown in Fig.~\ref{fig:diagrams}. As discussed in Ref.~\onlinecite{yWang2016}, in the forward
scattering limit the vertex corrections are of order $\lambda_m$, and can thus be neglected in the weak
coupling regime $\lambda_m \sim 0.15\text{--}0.25$ considered here. (Here, $\lambda_m$ measures the Fermi
surface average of the mass enhancement due to the {\ep} interaction, see Ref.~\onlinecite{Rademaker2016}.)
Note that the vertex correction is independent of the adiabatic parameter $\omega_\text{ph}/E_\text{F}$, in
contrast to the standard Migdal's approximation for $|\omega_\nu|/|\vq| \ll v_\text{F}$. (The vertex
correction is always proportional to $\lambda_m$ for either $|\omega_\nu|/|\vq| \ll v_\text{F}$ or
$|\omega_\nu|/|\vq| \gtrsim v_\text{F}$~\cite{yWang2016}, so our argument also applies for the
forward-focused {\ep} interaction.) There are alternative treatments that do not make use of Migdal's
approximation~\cite{Gorkov2016,Chubukov2016,Rosenstein2016} in the nonadiabatic regime for momentum
independent interaction $g(\vk,\vq) = g_0$. These approaches are beyond the scope of this work, which instead
focuses on a momentum dependent interaction. Furthermore, we calculate the dressed electron Green's function
(electron propagator) from the self-energy using the bare phonon Green's function (phonon propagator) [see
Fig.~\ref{fig:diagrams}(a)] and then insert this into the bubble diagram for the phonon self-energy [see
Fig.~\ref{fig:diagrams}(b)]. This approach is the so-called ``unrenormalized Migdal-Eliashberg''
scheme~\cite{Marsiglio1990}, where the phonon self-energy is not fed back into the electron self-energy
self-consistently. As we will show in section \ref{sec:screened_linewidth}, this treatment is justified when
one includes the Coulomb screening of the {\ep} interaction in the problem.

Adopting Nambu's 2-spinor scheme, the electron self-energy  $\hat{\Sigma}(\vk,\ri\omega_n) =
\ri\omega_n[1-Z(\vk,\ri\omega_n)]\hat{\tau}_0 +\chi(\vk,\ri\omega_n)\hat{\tau}_3 +\phi(\vk,\ri\omega_n)\hat{\tau}_1$
and the dressed electron Green's function $\hat{G}^{-1}(\vk,\ri\omega_n) =
\ri\omega_n\hat{\tau}_0 - \xi_\vk\hat{\tau}_3 - \hat{\Sigma}(\vk,\ri\omega_n)$ are matrices in Nambu space with
$\hat{\tau}_i$ being the Pauli matrices; $\omega_n = (2n+1) \pi/\beta$ are fermionic Matsubara frequencies with
$\beta = 1/T$ the inverse temperature ($k_\text{B}=1$); $Z(\vk,\ri\omega_n)$ and $\chi(\vk,\ri\omega_n)$
renormalize the single-particle mass and band dispersion, respectively; and $\phi(\vk,\ri\omega_n)$ is the
anomalous self-energy. The electron self-energy is self-consistently
calculated from the one-loop diagram in Fig.~\ref{fig:diagrams}(a) as follows
\begin{align}
  \hat{\Sigma}(\vk,\ri\omega_n)
  = & -\frac{1}{N\beta}\sum_{\vq,\nu} \Big[ |g(\vq)|^2 D_0(\vq,\ri\omega_\nu) \notag \\
    & \hat{\tau}_3\hat{G}(\vk-\vq,\ri\omega_n - \ri\omega_\nu)\hat{\tau}_3 \Big], \label{eq:elselfE}
\end{align}
where $D_0(\vq,\ri\omega_\nu) = -\frac{2\omega_\vq}{\omega^2_\vq + \omega_\nu^2}$ is the ``bare'' phonon
propagator.

Once we obtain the electron Green's function self-consistently, the polarization bubble in
Fig.~\ref{fig:diagrams}(b) is given by
\begin{align}
  P(\vq,\ri\omega_\nu)
  = & \frac{1}{N\beta}\sum_{\vk,n}\Tr \Big[ \hat{\tau}_3\hat{G}(\vk,\ri\omega_{n})\hat{\tau}_3 \notag\\
    & \hat{G}(\vk-\vq,\ri\omega_{n}-\ri\omega_\nu) \Big]\label{eq:ePoltr},
\end{align}
and $\Pi(\vq,\ri\omega_\nu) = |g(\vq)|^2 P(\vq,\ri\omega_\nu)$ is the phonon self-energy and
$\gamma(\vq,\omega) = -\Im \Pi(\vq,\ri\omega_\nu\to \omega+\ri\eta)$ is the phonon linewidth, which has been
analytically continued to the real frequency axis. To perform the analytic continuation we use the spectral
representation of the dressed Green's function
\begin{align}
  &\Im \Pi(\vq,\omega)
  = -|g(\vq)|^2\pi \int_{-\infty}^{\infty} d\omega' \Bigg\{
  \left[n_\text{F}(\omega' - \omega) - n_\text{F}(\omega') \right] \notag\\
  &\quad\quad  \frac{1}{N}\sum_\vk\Tr \left[ \hat{\tau}_3\hat{A}(\vk,\omega' - \omega)
        \hat{\tau}_3\hat{A}(\vk+\vq,\omega')  \right]
  \Bigg\},
\end{align}
where $n_\text{F}(x) =1/(e^{\beta x} + 1) $ is the Fermi-Dirac distribution function and
\begin{align}
  \hat{A}(\vk,\omega) = -\frac{1}{\pi}\Im \hat{G}(\vk,\omega+i\eta). \label{eq:AG}
\end{align}
$\hat{G}(\vk,\omega+i\eta)$ is obtained by the same iterative analytic continuation
method~\cite{Marsiglio1988} we used in Ref.~\onlinecite{Rademaker2016}.

Finally, we find the dressed phonon propagator using
\begin{align}
  D(\vq,\omega) &= \frac{2\omega_\text{ph}}{\omega^2-\omega_\text{ph}^2 + 2\ri\gamma(\vq,\omega)\omega_\text{ph}},
\end{align}
and phonon spectral function
\begin{align}
  B(\vq,\omega) &= -\frac{1}{\pi} \Im D(\vq,\omega).
\end{align}

In the numerical calculations, we solve the electron self-energy self-consistently on a $256\times 256$
$k$-grid. The convergence for the self-energy is reached if the difference of the self-energies from two
consecutive iterations is less than $10^{-3}\ut{meV}$. The small imaginary part included in the iterative
analytic continuation is $\eta = 3\ut{meV}$.

\section{Analytical Results for the Perfect Forward Scattering case} \label{sec:analytical}
We begin by examining the perfect forward scattering limit, where several analytical results can be obtained.
Here, we consider only the normal state in the low-temperature limit ($T_c < T \ll |\xi_\vk|$), because many
qualitative features of the phonon linewidth are already manifested there.

For a normal metal with a parabolic band $\xi_\vk = \frac{k^2}{2m}-E_\text{F}$, i.e., electron gas in
three-dimensions (3D), the analytical result of Eq.~(\ref{eq:ePoltr}) is the Lindhard
function~\cite{Fetter2003}. The corresponding result for electron gas in two-dimension (2D) is given in
Refs.~\onlinecite{Stern1967,Giuliani2005}. Without the {\ep} interaction, we can apply the 2D electron gas result
to our single band model, due to the small size of the Fermi pocket from the band dispersion $\xi_\vk =
-2t[\cos(k_x a)+\cos(k_y a)] - \mu \approx \frac{k^2}{2m^{*}}-E_\text{F}$, where $k=|\vk|=\sqrt{k_x^2 +
k_y^2}$, $m^{*}=\frac{1}{2t}$, $E_\text{F} = \frac{k^2_\text{F}}{2m^{*}}$, and $k_\text{F} =
\sqrt{4+\frac{\mu}{t}}$. This approximate band dispersion is exact at the band bottom and suitable for small
$k$. With this approximation, the imaginary part of the electron polarization without {\ep} interaction is
\begin{align}
  \Im P_0(\vq,\omega) =& -\frac{N_\text{F}}{\tilde{q}}
  \Big[ \Theta(1-\nu_{-}^2)\sqrt{1-\nu_{-}^2} \notag \\
  &- \Theta(1-\nu_{+}^2)\sqrt{1-\nu_{+}^2} \Big],
  \label{eq:ePol0}
\end{align}
where $\tilde{q}=|\vq|/k_\text{F}$, $\nu_{\pm} = \omega/(2E_\text{F} \tilde{q}) \pm \tilde{q}/2$, $N_\text{F}
= m^{*}/\pi$ is density of states of two spins, and the step-function $\Theta(x)=1$ for $x>0$ and
$\Theta(x)=0$ for $x<0$.

With the inclusion of the {\ep} interaction, the self-energy in Eq.~(\ref{eq:elselfE}) is nonzero but
diagonal in the normal state. In the perfect forward scattering limit $|g(\vq)|^2 = g_0^2 (N\delta_{\vq,0}) =
\lambda_m \omega_\text{ph}^2 (N\delta_{\vq,0})$, where $\lambda_m \equiv \la |g(\vq)|^2
\ra_\vq/\omega_\text{ph}^2 = g_0^2/\omega_\text{ph}^2$. The $(1,1)$-element of the self-energy is then given
by~\cite{Rademaker2016}
\begin{align}
  \Sigma(\vk, \ri\omega_n) &= \cfrac{a\omega^2_\text{ph}}{ \ri\omega_n-\xi_\vk-b\omega_\text{ph}
  - \cfrac{\omega^2_\text{ph}(1-b^2)}{\ri\omega_n-\xi_\vk+b\omega_\text{ph}} },
\end{align}
where $a=\lambda_m/\tanh\frac{\beta\omega_\text{ph}}{2}$ and $b=\tanh\frac{\beta\omega_\text{ph}}{2}
\tanh\frac{\beta\xi_\vk}{2}$. Using this self-energy and Dyson's equation, we find that at low temperatures
($T\ll |\xi_\vk|$ and $T\ll \omega_\text{ph}$), the dressed Green's function acquires a two-pole form
\begin{align}
  G(\vk, \ri\omega_n) = \frac{A_\text{M}}{\ri\omega_n-\xi_\vk^\text{M}} + \frac{A_\text{R}}{\ri\omega_n-\xi_\vk^\text{R}},
  \label{eq:GreenMRa}
\end{align}
where $A_\text{M,R} = (\sqrt{1+4\lambda_m}\pm 1)/(2\sqrt{1+4\lambda_m})$ and $\xi_\vk^\text{M,R} =
\xi_\vk+\frac{1}{2}\sgn(\xi_\vk) \omega_\text{ph}(1 \mp \sqrt{1+4\lambda_m})$. Here, ``M'' and ``R'' denote
the main and replica band, respectively. To simplify the calculation, we shift the two bands by the same
energy $-\frac{1}{2}\sgn(\xi_\vk) \omega_\text{ph}(1 - \sqrt{1+4\lambda_m})$ (which is small if $\lambda_m
\ll 1$), and the dressed Green's function becomes
\begin{align}
  G(\vk, \ri\omega_n) = \frac{A_\text{M}}{\ri\omega_n-\xi_\vk} + \frac{A_\text{R}}{\ri\omega_n-\xi_\vk^\text{R}},
  \label{eq:GreenMRb}
\end{align}
where the shifted $\xi_\vk^\text{R} = \xi_\vk+\sgn(\xi_\vk)\Delta\omega$ and $\Delta\omega =
\omega_\text{ph}\sqrt{1+4\lambda_m}$. Here, $A_\text{M} + A_\text{R} = 1$. Physically, Eq.
(\ref{eq:GreenMRb}) clearly indicates that the replica band exactly follows the dispersion of the main band,
and its energy offset from the main band is $+\Delta\omega$ ($-\Delta\omega$) for the part of the main band
above (below) the Fermi level.

\begin{figure}
  \centering
  \includegraphics[width=0.75\columnwidth]{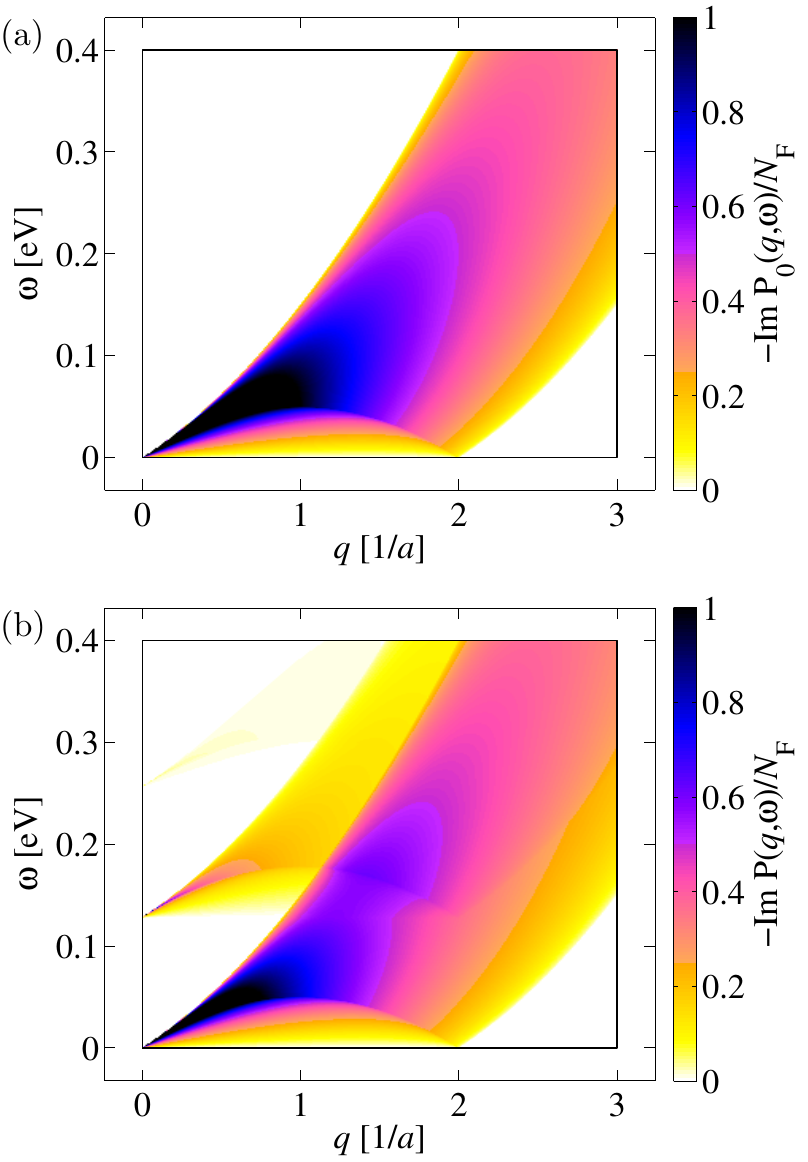}

\caption{Normalized imaginary part of the electron polarization $-\Im P(q,\omega)/N_\text{F}$ without {\ep}
interaction (a) and with forward scattering {\ep} interaction (b). $\lambda_m=0.16$ is used in panel (b). The
parabolic band approximation for FeSe/STO model $\xi_\vk\approx \frac{k^2}{2m^{*}} - E_\text{F}$ is assumed,
so $P(q=|\vq|,\omega)$ is isotropic in momentum space. $k_\text{F} \approx 1/a$, $v_\text{F} \approx 0.1
\ut{eV}\cdot\frac{a}{\hbar}$, $E_\text{F} \approx 0.05\ut{eV}$, and $\omega_\text{ph} = 0.1\ut{eV}$.}
  \label{fig:imPolT0}
\end{figure}

Using Eq.~(\ref{eq:GreenMRb}), the imaginary part of the electron polarization with the {\ep} interaction in
perfect forward scattering limit can be expressed in terms of the noninteracting electron polarization as
follows
\begin{align}
  &\Im P(\vq, \omega)
  = A_\text{M}^2 \Im P_0(\vq, \omega)
  \notag\\
  &+ 2A_\text{M} A_\text{R}
     \Im P_0\left(\vq, \omega-\sgn(\omega)\Delta\omega \right)\Theta\left(|\omega|-\Delta\omega\right)
  \notag\\
  &+ A_\text{R}^2
     \Im P_0\left(\vq, \omega-\sgn(\omega)2\Delta\omega\right)\Theta\left(|\omega|-2\Delta\omega\right).
  \label{eq:ePol}
\end{align}
Here, $\sgn(\omega)$ is the sign of $\omega$. Equation (\ref{eq:ePol}) is also a good approximation when the
coupling function $g(\vq)\propto \exp\left(-|\vq|/q_0 \right)$ has a sharp peak ($q_0\ll \pi/a$). Then, the
phonon linewidth is given by $\gamma(\vq,\omega) = -|g(\vq)|^2\Im P(\vq,\omega)$.

In Fig.~\ref{fig:imPolT0} we show $-\Im P({\bf q},\omega)$ calculated from Eq.~(\ref{eq:ePol0}) and from
Eq.~(\ref{eq:ePol}) in panel (a) and (b), respectively. Fig.~\ref{fig:imPolT0}(a) manifests the electron-hole
continuum for 2D electron gas at low temperature, while Fig.~\ref{fig:imPolT0}(b) shows multiple scattering
processes at low temperature corresponding to the three terms in Eq.~(\ref{eq:ePol}): one within the main
band for $|\omega|>0$ that represents the original electron-hole continuum, one between the main and replica
band for $|\omega| > \Delta\omega = \omega_\text{ph}\sqrt{1+4\lambda_m}$, and one within the replica band for
$|\omega| > 2\Delta\omega$, in a descending order of weights ($A_\text{M}^2$, $2A_\text{M} A_\text{R}$, and
$A_\text{R}^2$). As shown in Fig.~\ref{fig:imPolT0}(b), at the fixed frequency $\omega = \omega_\text{ph}$,
the magnitude of the imaginary part of the electron polarization has a sharp upturn at a finite momentum,
leading to a peak that slowly decreases at larger momentum. This qualitative feature persists in the full
numerical result in the next section. Note that since the coupling constant is a delta function, the phonon
linewidth $\gamma(\vq,\omega)$ is zero at all ${\bf q}$ values despite the fact that the polarization
$P(\vq,\omega)$ is nonzero.

\section{Numerical Results} \label{sec:numerical}
\begin{figure}
  \centering
  \includegraphics[width=1\columnwidth]{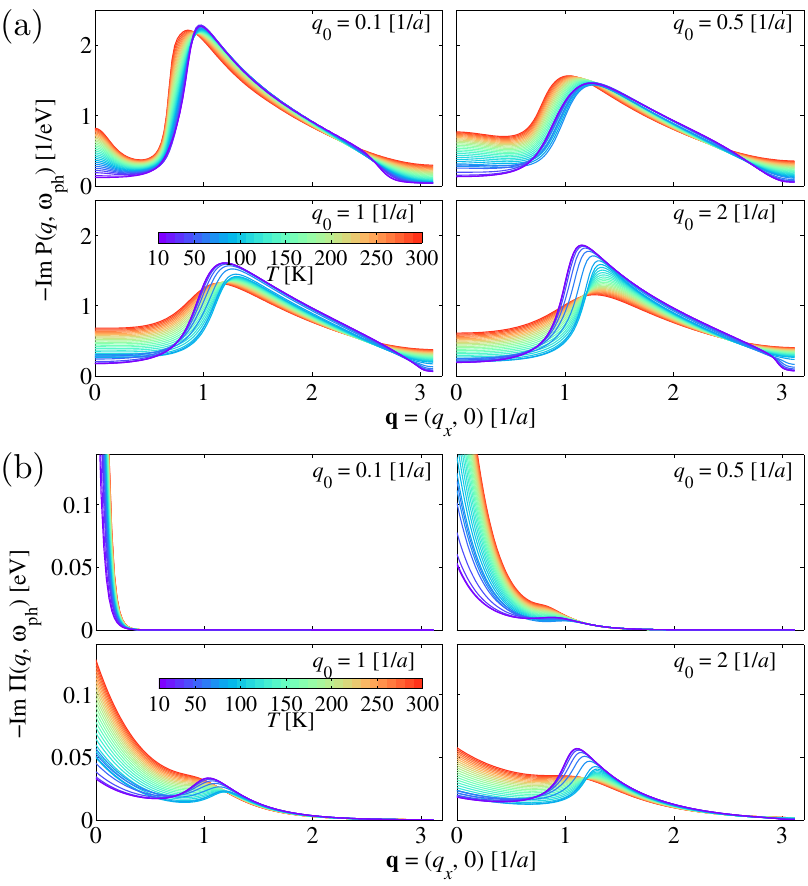}

\caption{(a) Momentum and temperature dependence of the imaginary part of the electron polarization $-\Im
P(q,\omega)$ for a fixed frequency $\omega = \omega_\text{ph}$ and various momentum width parameters
$q_0=0.1$, $0.5$, $1$, and $2$ in the {\ep} coupling function $g(\vq)\propto \exp(-|\vq|/q_0)$.  The double
Fermi-surface averaged coupling constant (defined in the text) is fixed at $\lambda = 0.8$. The colors (gray
scales) of lines represent low (blue) to high (red) temperatures. (b) Similar to (a) but for phonon linewidth
$\gamma(\vq,\omega) = -\Im \Pi(\vq,\omega) = -|g(\vq)|^2\Im P(\vq,\omega)$.}
  \label{fig:imPphT}
\end{figure}

We now turn to the polarization and phonon linewidth for the case of an {\ep} interaction with a small but
nonzero width in momentum space. Figure~\ref{fig:imPphT} shows the imaginary part of the electron
polarization $-\Im P({\bf q},\omega)$ and the phonon linewidth $\gamma_{\bf q}=-\Im\Pi({\bf
q},\omega_\text{ph})$ for various temperatures. Here, we have parameterized the total {\ep} coupling using
the double Fermi-surface averaged definition
\begin{align}
  \lambda = \frac{2}{\omega_\text{ph} \bar{N}_\text{F}N^2} \sum_{\vk,\vk'} |g(\vk-\vk')|^2  \delta(\xi_\vk)\delta(\xi_{\vk'}),
\label{eq:lambda}
\end{align}
where $\bar{N}_\text{F}$ is the density of states per spin and $N^{-2}\sum_{\vk,\vk'}
\delta(\xi_\vk)\delta(\xi_{\vk'}) = \bar{N}^2_\text{F}$. We have used this definition because the
$\vq$-averaged $\lambda_m = \la |g(\vq)|^2 \ra_\vq /\omega_\text{ph}^2 = g_0^2/\omega_\text{ph}^2$ equals the
mass enhancement factor $\left. -\Re \frac{\partial\Sigma(\omega)}{\partial\omega}\right|_{\omega=0}$ only in
the limit of perfect forward scattering, while $\lambda$ approximates the mass enhancement factor when the
{\ep} interaction is more uniform. The latter case occurs for the larger values of $q_0$ used in
Fig.~\ref{fig:imPphT}. In addition, $\lambda$ as defined in Eq. (\ref{eq:lambda}) does not depend on
temperature where as $\lambda_m$ does. Empirically, we find $\lambda_m \propto (q_0a)\lambda$ (see
Ref.~\onlinecite{yWang2016} for the proportionality constant), which can be used to approximately convert
between the two definitions. In Fig. \ref{fig:imPphT} we have set $\lambda=0.8$, which is equivalent to
$\lambda_m=0.16$ for $q_0=0.1/a$ and within the suitable range of values that simultaneously fit both high
$T_c$ value and the measured spectral weight of the replica bands~\cite{Rademaker2016}.

At low temperature and $\omega=\omega_\text{ph}$, the imaginary part of the polarization in
Fig.~\ref{fig:imPphT}(a) has a peak appearing at $|\vq|=\sqrt{2m^{*}\omega_\text{ph}}$, which is a feature of
the electron-hole continuum; with increasing temperature, the $-\Im P(\vq=0,\omega_\text{ph})$ increases, and
the rate of increase is faster for smaller values of $q_0$. The phonon linewidth, shown in
Fig.~\ref{fig:imPphT}(b), strongly peaks at $\vq=0$ for $q_0=0.1$ because the forward scattering coupling
$g(\vq)$ strongly suppresses the peak in the polarizability appearing at the finite $|\vq|$. As the value of
$q_0$ increases, however, the width of $g(\vq)$ begins to overlap with the peak in the polarization, and a
corresponding peak in the linewidth recovers at nonzero $\vq$. In this case, both the temperature and the
width of the coupling function $g(\vq)$ dictate the full $\vq$ dependence of the phonon linewidth. Thus, due
to its sensitivity to these parameters, the momentum dependence of the phonon linewidth can be used to
determine not only the overall strength of the {\ep} interaction but also the width of the coupling function.

To reproduce the replica bands observed in the ARPES experiments, the width of the {\ep} coupling must be
narrow in momentum space with $q_0 \approx 0.1/a\text{--}0.5/a$. Based on this observation, and the results
shown in Fig. \ref{fig:imPphT}, one might expect that the phonon linewidth in the vicinity of $\vq = 0$
should be very large. In turn, the real part of the phonon self-energy will also develop significant Kohn
anomaly, leading to an instability of the lattice. It turns out that the Coulomb interaction will prevent
this from occurring, as the divergence in the Coulomb interaction at $\vq = 0$ effectively blocks the
long-wavelength instability. We will discuss this issue in the next section.

\section{Undressing of the phonon linewidth due to Coulombic Screening}\label{sec:screened_linewidth}
In this section we examine the effects of Coulomb screening by the FeSe electrons on the {\ep} vertex and the
phonon linewidth. Fig.~\ref{fig:diagrams}(c) shows the diagramatic expansion of the {\it screened} {\ep}
vertex evaluated at the level of the random phase approximation. The screened vertex is
\begin{align}
    \bar{g}(\vq,\ri\omega_\nu)
  &= g(\vq) + g(\vq)\left[-V_C(\vq)\chi_0(\vq,\ri\omega_\nu)\right]          \notag \\
  &\quad + g(\vq)\left[-V_C(\vq)\chi_0(\vq,\ri\omega_\nu)\right]^2+\dots \notag \\
  &= \frac{g(\vq)}{1+V_C(\vq)\chi_0(\vq,\ri\omega_\nu)},
\end{align}
where $\chi_0(\vq,\ri\omega_\nu) = -P(\vq,\ri\omega_\nu)$ is the charge susceptibility and $V_C(\vq)$ is the
Fourier transform of the Coulomb potential. In the contiuum limit, $V_C(\vq) = \frac{4\pi e^2}{|\vq|^2}$ in
three dimensions (3D) and $V_C(\vq) = \frac{2\pi e^2}{|\vq|}$ in two dimensions (2D). The corresponding
phonon self-energy is obtained by replacing the vertex function with the screened vertex with
\begin{align}
     \Pi(\vq,\ri\omega_\nu)
  &= g(\vq)\left[\bar{g}(\vq,\ri\omega_\nu)\right]^* \left[-\chi_0(\vq,\ri\omega_\nu) \right] \notag \\
  &= \frac{-|g(\vq)|^2 \chi_0(\vq,\ri\omega_\nu) }{1+V_C(\vq)\chi_0(\vq,\ri\omega_\nu)},
\end{align}
where we have assumed $V_C(\vq)\chi_0(\vq,\ri\omega_\nu)$ is real.

Here, we are interested in the case of an FeSe monolayer located a distance $h$ above the oxide substrate. We
place the FeSe electrons at $z = 0$ and the ions in the termination layer of the substrate at $z = -h$. For
this geometry, we introduce an anisotropic Coulomb potential~\cite{Aristov2006}
\begin{align}
  V_C(q,q_z) &= \frac{4\pi e^2}{\epsilon_a q^2 + \epsilon_c q^2_z},
\end{align}
where $q = \sqrt{q_x^2 + q_y^2}$ is the momentum transfer a plane parallel to the FeSe monolayer, and
$\epsilon_a$ and $\epsilon_c$ are the zero-frequency dielectric constants parallel and perpendicular to the
plane. By inverse Fourier transform, the real space formula is
\begin{align}
  V_C(x,y,z) &= \frac{e^2}{\sqrt{\epsilon_a\epsilon_c}} \frac{1}{\sqrt{r^2 + \bar{z}^2}},
\end{align}
where $r^2 = x^2 + y^2$ and $\bar{z} = (\epsilon_a/\epsilon_c)z^2$. After performing the 2D
fourier transforming for the in-plane coordinates we arrive at
\begin{align}
  V_C(q,z) &= \frac{2\pi e^2}{\sqrt{\epsilon_a\epsilon_c}}\frac{e^{-q|\bar{z}|}}{q}.
\end{align}
To compute the screened {\ep} interaction, we must use the interaction at $z = 0$ for the Coulomb potential
since the particle-hole pairs are created in the FeSe layer. Putting this all together, the phonon linewidth
is given by
\begin{align}
  \gamma(\vq,\omega) &= \frac{\omega_\vq}{\omega_\text{ph}}
    \Im \frac{|g(\vq)|^2\chi_0(\vq,\omega)}{1+V_C(q,z=0)\chi_0(\vq,\omega)/a^2}
\label{eq:screened_gamma}
\end{align}
where we define the ``unscreened" phonon energy as $\omega_\vq = \sqrt{\omega_\text{ph}^2 + [\Re
\Pi(\vq,\omega)]^2} - \Re \Pi(\vq,\omega)$.

\begin{figure}
  \centering
  \includegraphics[width=\columnwidth]{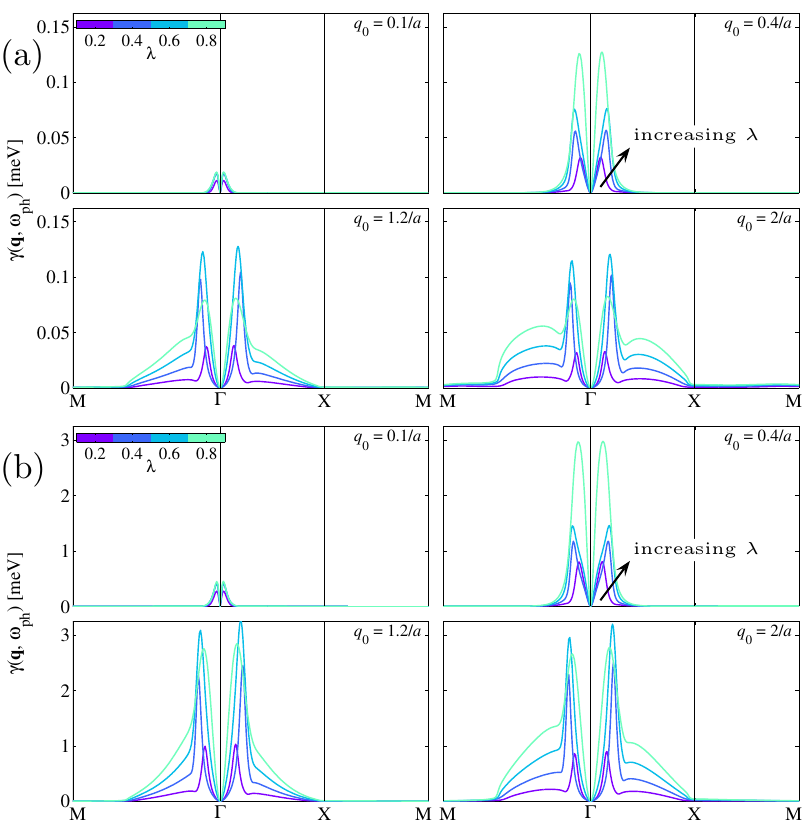}

\caption{The phonon linewidth $\gamma(\vq,\omega=\omega_\text{ph})$ along a high symmetry path
$M$-$\Gamma$-$X$-$M$ at $T = 30$~K. Results are shown for (a) $\epsilon_a = 1 = \epsilon_c$ and (b)
$\epsilon_a = 25$, $\epsilon_c = 1$. The line color (gray scale) encodes the different values of
$\lambda=0.2$, $0.4$, $0.6$, and $0.8$, as indicated by the color bar.}
  \label{fig:screened_gamma}
\end{figure}

We evaluated Eq.~(\ref{eq:screened_gamma}) for several values of $q_0$ and $\lambda$, and the results are
shown in Fig.~\ref{fig:screened_gamma}. Since the exact values of the dielectric constants are not known for
the FeSe interface systems, we show results for $\epsilon_a = \epsilon_c = 1$ in
Fig.~\ref{fig:screened_gamma}(a) and $\epsilon_a = 25$, $\epsilon_c = 1$ in Fig.~\ref{fig:screened_gamma}(b).
Note that the latter values are close to the estimates obtained by Kuli{\'c} and Dolgov
(Ref.~\onlinecite{Kulic2017}) in the limit of perfect forward scattering. In both cases, we find that the
phonon linewidth is dramatically suppressed once Coulomb screening is included; however, as the values of
$\epsilon_a$ and $\epsilon_c$ are increased, the magnitude of the linewidth increases. These results indicate
that the long-range Coulomb interaction prevents the formation of a competing charge ordering at
long-wavelengths, which is consistent with the notion that extend Coulomb interactions can suppress
insulating behavior~\cite{Poilblanc1997}. Our results also show that this effect will be somewhat senstive to
the dielectric properties of the interface, which may offer a means to tune these properties. Finally, the
undressing of the phonon linewidth observed here also provides a rationale for adopting an unrenormalized
Migdal-Eliashberg scheme, where the phonon self-energy is not fed back into the electron self-energy in a
self-consistent manner. In this case, the calculated phonon self-energy is small, justifying the use of the
bare phonon propagator in the electron self-energy diagrams.

Comparing our results to the recent RHEELS measurements by Zhang \textit{et al.}~\cite{Zhang2016}, we find
that once the Coulomb screening is included, the computed linewidths are much smaller than those inferred
experimentally. Moreover, in the experimental data, the linewidth is finite at $\Gamma$ point and maximal
around $X$ point. Our calculated linewidth is exactly zero at $\Gamma$ point because the screening from the
Coulomb potential diverges at $\vq = 0$. However, we have not considered any impurity potential in our
calculation, or other sources of broadening in the electron Green's function, and, subsequently, the phonon
linewidth once the charge susceptibility $\chi_0$ is computed. Regardless, $X$ point is not the maximal point
for the linewidth in any of our calculation results. This discrepancy could also be due to the limitation of
our single band model. The real system is multiband in nature and also shows strong magnetic fluctuations.

\section{Summary and Conclusions}\label{sec:summary}
In this paper, we have calculated the phonon linewidth, i.e., the imaginary part of the phonon self-energy in
an unrenormalized Migdal-Eliashberg scheme in the weak to intermediate coupling regime for strong forward
scattering {\ep} interaction. Such an {\ep} interaction dresses the electron propagator by simply creating
the replica bands and shuffles the electron-hole continuum of 2D electron gas into three similar parts with
descending weights beginning at $|\omega|>0$, $|\omega|>\Delta\omega$, and $|\omega|>2\Delta\omega$. If we do
not include Coulomb screening, the phonon linewidth is a simple product of coupling function $|g(\vq)|^2$
with a forward scattering peak around $\vq=0$ and the electron polarization with a very similar momentum
structure of the electron-hole continuum of 2D electron gas. Depending on the peak width $q_0$ of the {\ep}
coupling constant $g(\vq)$ and the peak of electron polarization at $|\vq|=\sqrt{2m^{*}\omega_\text{ph}}$, we
find the linewidth $\gamma(\vq,\omega_\text{ph})$ has a maximum value at $\vq=0$ or $|\vq| \approx
\sqrt{2m^{*}\omega_\text{ph}}$ at low temperature, and the linewidth is broad at $\vq = 0$. Even if the
latter happens, since the linewidth for small $|\vq|$ tends to increase with temperature, the maximum may
shift back to $\vq=0$ at high temperature. The momentum resolved phonon spectral function at $\omega \approx
\omega_\text{ph}$ can be understood in the same picture.

The broad linewidths at $\vq = 0$ would normally indicate an instability to a charge-ordered phase at long
wavelengths. However, once the long-range Coulomb interaction screens the {\ep} interaction we find that the
phonons are undressed. Here, the anomalous broadening at $\vq = 0$ is suppressed by the divergence in the
Coulomb interaction at $\vq = 0$ while the total phonon linewidth is reduced throughout the Brillioun zone.
In this case, a small peak remains at nonzero momentum transfers; however, the magnitude of this peak is much
smaller than the linewidths measured by HREELS~\cite{Zhang2016}. Our results suggest that the broadening of
the {\STO} phonons (with a maximum at $X$ point) observed by Zhang \textit{et al.} are not due to the
forward-focused {\ep} coupling inferred from the ARPES measurements and are likely from some other source. To
resolve the forward-focused {\ep} interaction, the HREELS experiments should focus on smaller values of
$\vq$, which will be challenging given the large background signal at $\vq = 0$.

\begin{acknowledgments}
We thank T.~Berlijn, T.~P.~Devereaux, M.~L.~Kuli{\'c}, E.~W.~Plummer, and D.~J.~Scalapino for useful
discussions. L.~R. acknowledges funding from the Dutch Science Foundation (NWO) via a Rubicon Fellowship and
by the National Science Foundation under Grant No. PHY11-25915 and Grant No. NSF-KITP-17-019. S.~J. is funded
by the University of Tennessee's Office of Research and Engagement's Organized Research Unit program.  Y.~W.
and E.~D. are supported by the U.S. Department of Energy, Office of Basic Energy Sciences, Materials Sciences
and Engineering Division. This research used computational resources supported by the University of Tennessee
and Oak Ridge National Laboratory's Joint Institute for Computational Sciences.
\end{acknowledgments}

\end{document}